\documentclass[12pt,twocolumn,a4paper]{article}

\usepackage{graphicx}

\def\eep{\mbox{$(e,e^{\prime}p)$}}
\def\eepn{\mbox{$(e,e^{\prime}pN)$}}
\def\O16{$^{16}$O}

\def\emiss{$E_m$}
\def\pmiss{$p_m$}
\def\He4{$^4$He}
\def\omegaq{${\omega},q$}
\def\gammapq{$\gamma_{pq}$}
\def\mevc{MeV/$c$}
\def\degree{$^\circ$}

\begin{document}

\onecolumn
\begin{center}
{\bf{\Large{\O16\eep\ reaction at large missing energy}}}
\vspace{0.5cm}

M. Iodice$^1$, E. Cisbani$^2$, R. De Leo$^3$, S. Frullani$^4$, F. Garibaldi$^4$, D.L.~Groep$^5$,  \\
W.H.A.~Hesselink$^{4,5}$, E.~Jans$^4$, L.~Lapik\'{a}s$^4$, C.J.G.~Onderwater$^{4,5}$, \\
R.~Perrino$^6$, J.~Ryckebusch$^{7}$, R.~Starink$^{4,5}$, G.M.~Urciuoli$^8$  \\
\vspace{0.5cm}
{\footnotesize \it
$^{1}$ INFN, Sezione Roma Tre, Via della Vasca Navale, 84, I-00146 Roma, Italy \\
$^{2}$ INFN, Gruppo Collegato Sanit\`a and Istituto Superiore di Sanit\`a,\\
$^{3}$ University and INFN Bari, Via Amendola 173, I-70126 Bary, Italy \\
       Viale Regina Elena, 299, I-00161 Roma, Italy \\
$^{4}$ NIKHEF, P.O.~Box~41882, 1009~DB~Amsterdam, The~Netherlands \\
$^{5}$ Vrije~Universiteit~Amsterdam, de~Boelelaan~1081, 1081~HV~Amsterdam, The~Netherlands \\
$^{6}$ INFN Sezione di Lecce, via Arnesano, I-73100 Lecce, Italy \\
$^{7}$ Department of Subatomic and Radiation Physics, Ghent University, Proeftuinstraat 86,
       B-9000 Gent, Belgium  \\
$^{8}$ INFN, Sezione Roma, Piazzale Aldo Moro, 2, I-00185 Roma, Italy \\
}
\vspace{0.5cm}
\today \\
\end{center}

{\footnotesize
{\bf Abstract}

We investigate the origin of the strength at large missing energies in electron-induced proton knockout reactions. For that purpose the reaction \O16\eep\ was studied at a central value $\omega$ = 215 MeV of the energy transfer, and two values of the momentum transfer: $q = 300, 400$ \mevc, corresponding to the ``dip region''.
Differential cross sections were determined in a large range of
missing energy (\emiss=0--140 MeV) and proton emission angle (\gammapq=0--110\degree),
and compared to predictions of a model that includes
nucleon-nucleon short-range correlations and two-body currents.
It is observed that, in the kinematic domain covered by this experiment,
the largest contribution to the cross section stems from two-body currents,
while short-range correlations contribute a significant fraction.

\vspace{0.5cm}
\noindent
{\it PACS:} 21.10.Pc; 24.10.-i; 25.30.Fj
% 21.10.Pc 	Single-particle levels and strength functions
% 24.10.-i 	Nuclear reaction models and methods
% 25.30.Fj 	Inelastic electron scattering to continuum

\noindent
{\it Keywords:} Electron-induced knockout reactions; Short-range correlations; Two-body currents
}
%\newpage
\twocolumn

% In the missing-energy range \emiss\ = 0-60 MeV the cross sections for electron-induced proton knockout A\eep\ are dominated by direct knockout processes, where the proton that is detected is the one that was hit by the transfered virtual photon (Impulse Approximation). The recoiling (A-1) nucleus is left in an (excited) state, of which most properties are well described by mean-field theory. At larger \emiss\ still an appreciable strength is measured (see for example \cite{Marchand,lee01,Benmokhtar} and \cite{Fissum} and references therein) that can not be explained by direct knock-out processes. Here, two-nucleon and multi-nucleon knockout mechanisms must play a role that lead to one proton that is detected and other nucleon(s) that go unobserved.
Electron-induced proton knockout  A\eep\ reactions, at low values of missing energy \emiss\ and missing momentum \pmiss, are dominated by direct processes, where the detected proton is the one that was hit by the transferred virtual photon (Impulse Approximation), while the recoiling (A-1) nucleus is left in an (excited) state, of which most properties are well described by mean-field theory. At larger \emiss\ and \pmiss, other reaction mechanisms, as two- and multi-nucleon knockout, start to play a role. Here the undetected (A-1) system could consist of a residual nucleus and one or more nucleons that were correlated with the hit nucleon and that have been knocked out  in the reaction. These mechanisms should be dominant when the values of missing energy are higher than those expected from mean-field theory for the deepest bound state and where still an appreciable strength is measured (see for example \cite{Marchand,lee01,Benmokhtar,Fissum} and references therein).
The aim of the present paper is to examine the origin of this excess strength in terms of short-range correlations (SRC) and two-body currents. For that purpose two channels can be studied:
exclusive two-nucleon knockout or semi-inclusive one-nucleon knockout.

In the two-nucleon knockout \eepn\ reaction, SRC in nuclei are probed directly
by measuring the cross section for transitions to selected states
at small excitation energy in the residual nucleus {\cite{ond98,gro99,sta00,niy03}}.
The single nucleon-knockout \eep\ reaction
can provide information on  the nucleon spectral function 
at large energy and momentum, which is sensitive to the nucleon-nucleon
interaction at short distance (cf. Ref. \cite{jan00} and \cite{rohe04}).
However, in both cases other competing processes contribute to the cross section.
In particular, two-body currents, which include meson-exchange
currents (MEC) and intermediate ${\Delta}$-excitation with a subsequent
${\Delta}N \rightarrow NN$ interaction, are known to make a substantial
contribution to both the cross section of the exclusive \eepn\ reaction and
the semi-inclusive \eep\ reaction.
Hence, for these reactions a decomposition of the cross section
into contributions from one-body and two-body hadronic currents
can only be made by comparing the data with calculated cross sections 
that include both processes.

\begin{figure*}[!ht]
\begin{center}
\includegraphics[height=12.5cm, width=12.5cm]{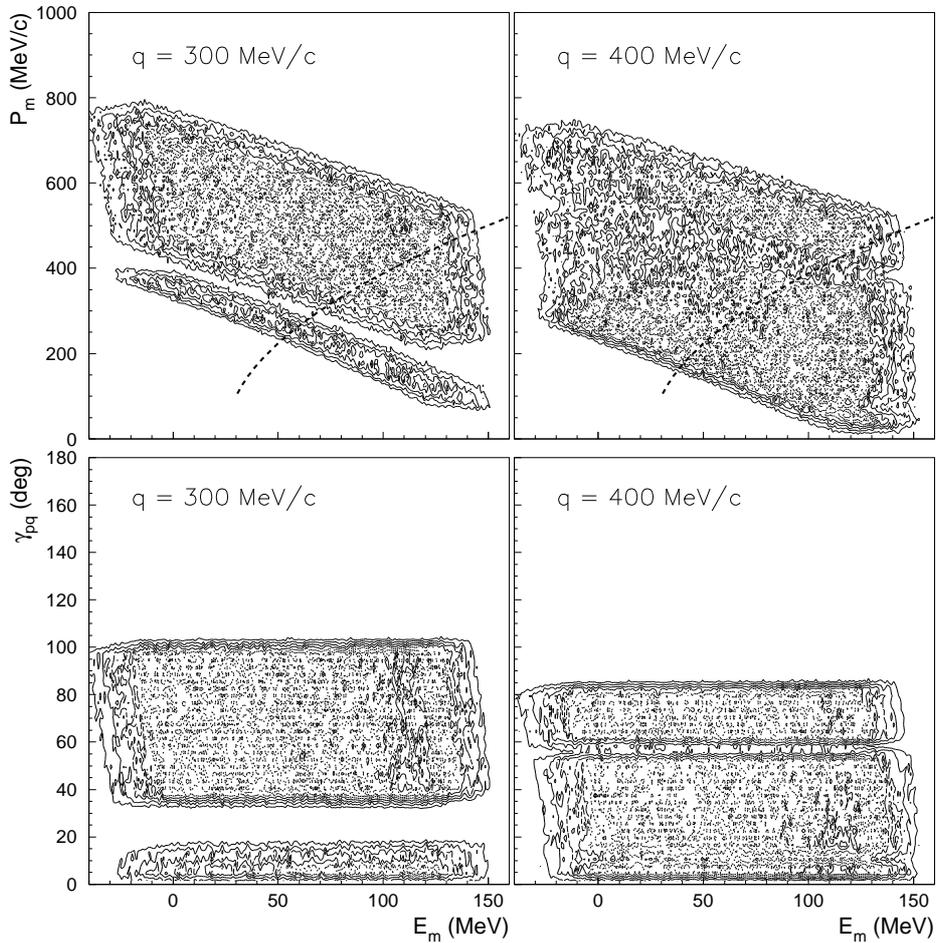}
\caption{\footnotesize Contours of the detection volumes in the (\emiss, \pmiss) phase space (top
 panels) and (\emiss, \gammapq) plane (bottom panels) for the kinematics
with $q$ = 300 \mevc\ (left) and 400 \mevc\ (right).
The dashed curves show the position of the ridge according to Eq.~(\ref{eq:Em-ridge}).}
\label{phase_space}
\end{center}
\end{figure*}

The semi-inclusive \eep\ data have to be compared
to cross sections calculated for a large domain in missing energy \emiss\ and
missing momentum \pmiss, because the relative contribution of either of the
two processes to this reaction depends on these kinematic variables.
Furthermore, in choosing the (\emiss, \pmiss) domain
for which this comparison is made,
one has to take into account that in the process of interest
two particles are emitted and only one is detected. Thereby,
\emiss\ and \pmiss\ are largely determined by
the kinetic energy and momentum of the undetected nucleon.
Neglecting the momentum and intrinsic excitation of the recoiling (A-2) nucleus,
the non-relativistic relation between these quantities reads:
\begin{equation}
E_m = E_{thr} + {\frac{A-2}{A-1}}\;{\frac{p^2_m}{2M}},
\label{eq:Em-ridge}
\end{equation}
where $M$ is the nucleon mass, the threshold energy for two-nucleon knockout
$E_{thr}$ = $E_A$ - $E_{(A-2)}$ is the difference between the binding energies of the A and (A-2) nucleon systems,
and the factor ${\frac{p^2_m}{2M}}$ in the second term is the kinetic energy  
of the undetected nucleon. This term accounts for the excitation energy of
the (A-1) system \cite{cio96}.
Eq.~(\ref{eq:Em-ridge}) indicates that the largest cross sections are expected
along a parabolic curve in the (\emiss, \pmiss) plane, shown in Fig.~\ref{phase_space}.
The pair-momentum distribution inside the nucleus, final-state interactions
and differences in intrinsic excitation of the (A-2) nucleon system cause a smearing
of this reaction strength.
Hence, one may expect a broad ridge in the (\emiss, \pmiss) plane, the top location
of which is represented by Eq.~(\ref{eq:Em-ridge}).
These features for semi-inclusive \eep\ reactions at kinematic conditions
that are characteristic for two-nucleon knockout, are confirmed by the (\emiss, \pmiss)
spectra measured previously for the \He4\eep\ reaction at energy and momentum
transfer: (\omegaq) = (215 MeV, 401 \mevc) \cite{lee01}.
The values of (\omegaq) chosen for the \O16\eep\ study discussed in
this Letter, are about the same as those for the \He4\ experiment. This allows a comparison between
both data sets, which is of interest because, in a mean-field picture, in \He4\ the nucleons are knocked out of the $1s$ shell,
whereas in \O16\ knockout of nucleons from the $1p$ shell is the dominant process.
A detailed comparison between the data for \O16\ and  \He4\ will allow to study the mass
dependence of short-range correlations in nuclei.

The experiment was performed in the EMIN (external target) hall of the Amsterdam
Pulse Stretcher (AmPS) facility~\cite{wit93} at NIKHEF.
Electrons extracted from AmPS had an energy of 575 MeV.
The target used in this experiment was a single foil waterfall target \cite{gar92}. 
A thin film of water was formed due to surface tension and 
adherence to two metal bars positioned below a thin slit, through which the water 
was pumped from a reservoir. 
The scattered electrons and emitted protons 
were detected in the QDQ high-resolution magnetic spectrometer and
in the large-acceptance scintillator detector
HADRON3~\cite{pel99}, respectively. 
The latter detector consists of
two hodoscope layers (H1 and H2) and six energy determining layers 
(L1-L6) of plastic scintillator slabs.
It features an angular acceptance of $\pm$14\degree\ and a
nominal energy acceptance of 50-225 MeV. For the present experiment the low-energy threshold was raised to 65 MeV by installation of a 5 mm thick Pb-shield in front of the detector.
Thus, the detector could be placed
at angles as forward as 30\degree\ without being swamped by low-energy protons,
while simultaneously allowing the detection of protons along the momentum transfer $\mathbf{q}$. 
The angular resolution was 0.52\degree\ both in and out of the reaction plane.

Before starting data taking the energy-response of HADRON3 was calibrated
using the continuous energy-spectrum of the detected protons (singles events).
The high voltages of all photo-multipliers were set such that the ADC value corresponding
to the maximum energy-loss of the protons in a specific layer was about 75\% of the range.
The kinetic energy of the protons was determined from the amount of
light measured in the layer in which the particle was stopped. This
amount of light and that measured
in the preceding layer were used for particle identification \cite{pel99}.
The arrival time of the protons in the HADRON3 detector was extracted from 
the signals in the first energy-determining layer. A time resolution of 700 ps (FWHM) was 
achieved for coincidences between electrons and protons. 

During several dedicated runs the cross-section for elastic electron scattering  
off \O16\ was measured, using events that were recorded with the QDQ magnetic
spectrometer. From these data the thickness of the waterfall
target was determined using the
known cross section for elastic electron scattering
off \O16\ \cite{elasOxy}. The measured thickness was $173\pm4$ mg/cm$^2$.
The error is completely dominated by the systematic experimental uncertainties.
Throughout the experiment the target thickness was 
monitored by comparing the singles rates in the QDD spectrometer and 
HADRON3 detectors. 
These data indicate that over the complete experiment variations in the 
thickness of the water film were within a range of $\pm 3\%$. 
The variations are known with an accuracy of much better than 1\% 
and accounted for in the normalization.

Data for the \O16\eep\ reaction were taken at an average transferred energy $\omega$  =  210~MeV
and at two values of the average transferred momentum $q$: 300 \mevc\ (kinematics I) and
400~\mevc\ (kinematics II).
These values of  (\omegaq) correspond to
%$\frac{Q^2}{2M {\omega}} \approx $ 0.2 and 0.4, respectively. They are characteristic for
the so called ``dip region'', the domain between the cross section maxima for quasi-elastic scattering and $\Delta$-production.

For each of the kinematics I and II electron-proton coincidences were measured at four angular settings $\theta_p$
of the HADRON3 detector. They correspond to the following ranges in the proton emission-angle 
\gammapq $=\theta_p - \theta_q$ with respect to the direction $\theta_q$ of the momentum-transfer vector $\mathbf{q}$:
0\degree$\leq$\gammapq$\leq$102\degree\ for kinematics I ($q$=300 \mevc, $\theta_q$=33.6\degree) and
0\degree$\leq$\gammapq$\leq$85\degree\ for kinematics II ($q$=400 \mevc, $\theta_q$=39.3\degree).
In the lower part of Fig.~\ref{phase_space} the contours of the detection volumes, expressed in  \emiss\
and \gammapq, are displayed for the two values of the transferred momenta.
The upper part of Fig.~\ref{phase_space} shows the corresponding phase-spaces in missing energy and missing momentum.
They span the ranges $-20\leq E_m \leq 150$ MeV and $100\leq p_m \leq 800$ \mevc, respectively.

The data were corrected for radiative effects using the formalism developed  by Mo and Tsai
\cite{mo69}. Effects due to internal and external Bremsstrahlung as well as ionisation
were taken into account.
The electromagnetic character of the radiative effects allows
a precise calculation of the transfer of reaction strength from small to large \emiss\ values,
while \pmiss\ can either increase or decrease.
The unfolding of the radiative effects in the measured cross sections
as a function of \emiss\ and \pmiss\ was conducted as follows.
Starting at the lowest \emiss\ bin, the cross section in each \pmiss\ bin
corresponding to this \emiss\ value was corrected for the loss of events.
Next the radiative strength stemming
from the first \emiss\ bin was subtracted from all other bins in (\emiss, \pmiss).
This procedure was repeated for the (corrected) strength of the second \emiss\ bin and
subsequently for all other bins until the complete (\emiss, \pmiss) spectrum was unfolded.
Sizable contributions to the cross sections may stem from radiative effects
originating from (\emiss, \pmiss) regions that are not covered by the kinematic acceptance
of the experiment. Especially the strength originating from the valence shells
in the \pmiss\ range 50-150 \mevc\ can contribute appreciably.
The radiative effects associated with knockout from
the $1p_{1/2}$ and $1p_{3/2}$ single-particle levels in \O16\ were estimated using a fit
to their momentum distributions determined from high-resolution data previously measured at NIKHEF \cite{leusch94}. To minimize the influence of (\emiss, \pmiss)
bins with a very small number of events and consequently a small statistical accuracy,
for each \emiss\ bin fits as function of \pmiss\ were made to the data.
Then, the results of the fits were used in the radiative unfolding procedure.

\begin{figure}[ht]
\includegraphics[width=7cm]{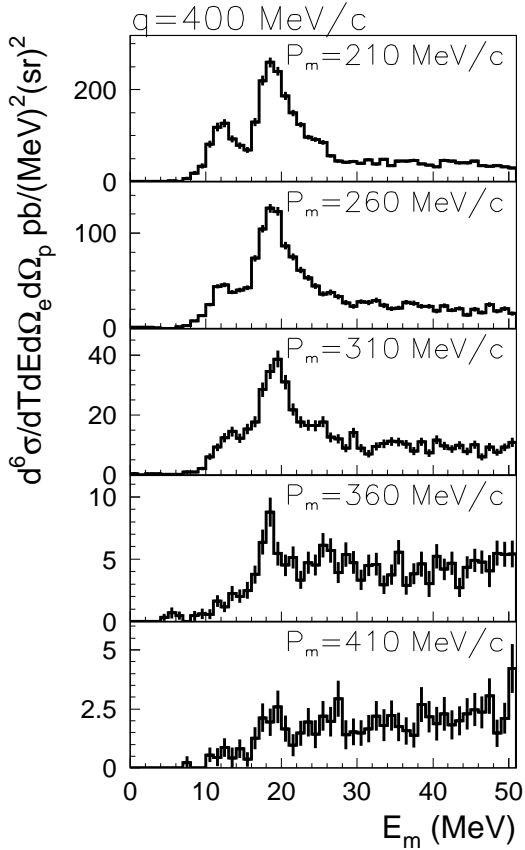}
\caption[radtail]{\footnotesize Missing-energy spectra for the \O16\eep\ reaction
  at a momentum transfer $q = 400$ \mevc, for various values of \pmiss.
  The error bars denote statistical uncertainty only.}
\label{lowem}
\end{figure}

\begin{figure*}[!ht]
\begin{center}
\includegraphics[height=12.5cm, width=12.5cm]{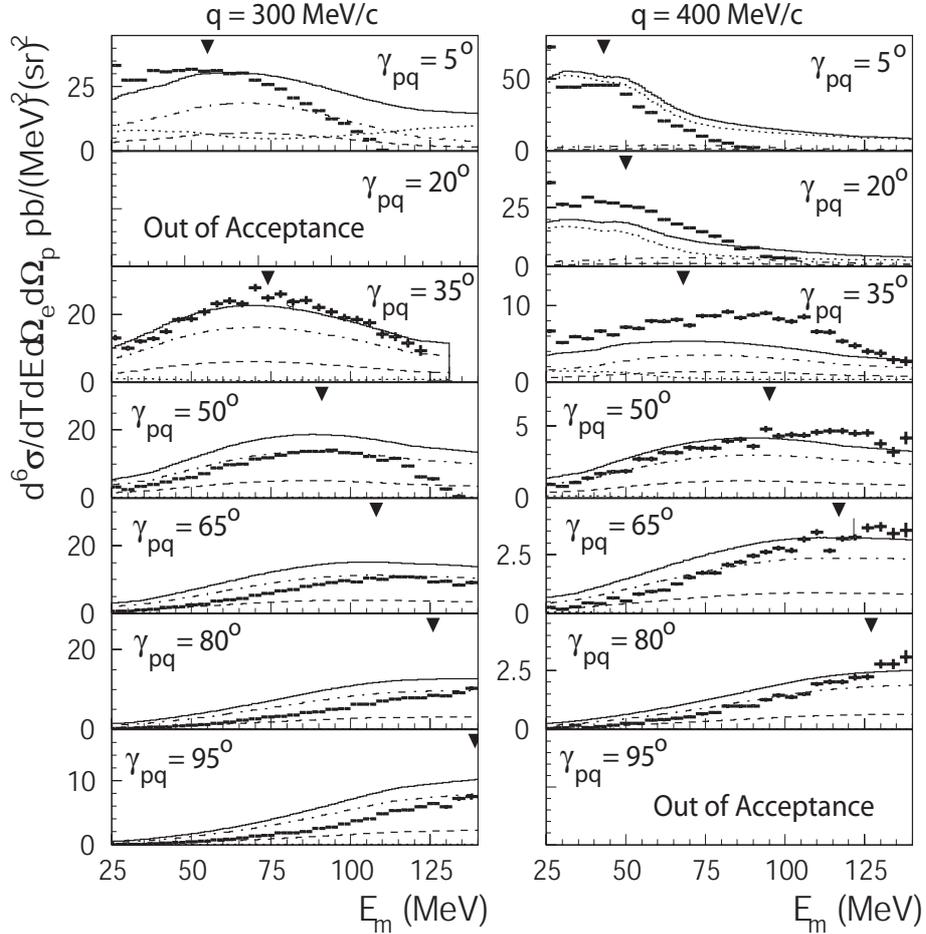}
\caption{\footnotesize Missing-energy spectra for the \O16\eep\ reaction
at a momentum transfer $q = 300$ \mevc\ (left panel) \, and \, 400 \mevc\ \, (right panel).
The dotted curves correspond to calculated cross sections for knockout
of a proton from the $s$-shell. The dashed (dot-dash) curves represent
the results from calculations of two-nucleon knockout due to one-body
(two-body) hadronic currents. The solid curve is the coherent sum of all contributions.
The black triangles show the position where the maximum strength from correlated pairs is expected
according to Eq.~(\ref{eq:Em-ridge}).}
\label{highem}
\end{center}
\end{figure*}

From the data taken at each of the two combinations
of (\omegaq) the measured cross sections,
differentiated with respect to the electron and proton solid angles (d$\Omega_p$, d$\Omega_e$) and to the proton kinetic energy (d$T$) and electron scattering energy (d$E$), are shown in Figs. \ref{lowem} and \ref{highem}.
The total systematic error in the cross section amounts to 7\%.
It results from the quadratic sum of 5\% uncertainty in the radiative unfolding
procedure including extrapolation of the continuum strength outside the         
measured region, and a 5\% systematic experimental error. The latter results
mainly from the uncertainties in the HADRON3 detection efficiency (3\%) and
in the target thickness determination (2.5\%).

Fig.~\ref{lowem} shows the cross sections measured at a momentum transfer $q = 400$ \mevc.
They are displayed as a function of \emiss\ in the range 0--50 MeV, at five central values of
the missing momentum.
It is clear that the cross section at small \emiss\ (10 $ \leq E_m \leq $ 20 MeV),
corresponding to knockout of a proton from the $1p$ shell and leaving the residual nucleus in
a state with small excitation energy, decreases rapidly at
increasing missing momentum.
This trend is characteristic for the proton momentum distribution in a nucleus
as calculated in a mean field approach. A similar observation has been made for the
\He4\eep\ reaction (cf. Ref. \cite{lee01}).
On the contrary, above \emiss$\approx$25 MeV the \pmiss\ dependence of the cross section is softer.
In this domain knockout of two or more nucleons gradually becomes the dominating reaction mechanism (see Fig.~\ref{lowem}).
According to Eq.~(\ref{eq:Em-ridge}), the missing energy \emiss\ and missing momentum \pmiss\ are correlated
in a two-nucleon knockout reaction in which only one nucleon is detected.
Indeed, the maximum of the continuum cross section in Fig.~\ref{lowem} is seen to shift toward higher \emiss\ with increasing \pmiss.

In Fig.~\ref{highem} the six-fold differential cross sections for the \O16\eep\ reaction
measured as a function of the missing energy at 
$q = 300$ \mevc\ (left) and 400 \mevc\ (right), are presented for seven  bins in
\gammapq, together with calculated cross sections.
This part of the \emiss\ spectrum contains the information on short-range correlations and
other processes that contribute to two-nucleon knockout.
Two domains do not contain data. They are out of the acceptance covered in the present experiment (see Fig.~\ref{phase_space}).

The theoretical cross sections for two-nucleon knockout, also shown in Fig.~\ref{highem},
were evaluated in an unfactorized
framework based on the assumption mentioned above, i.e. that in a semi-inclusive
\eep\ reaction at large \emiss\ two nucleons are emitted, of which one is detected
\cite{jan00}. They include contributions of one-body as well as two-body hadronic currents,
and can be considered as an extension of the two-nucleon knockout model for transitions to 
the ground state and states at low excitation energy in the \O16\eepn\ reaction
($N$ is either a proton or a neutron) to energies beyond the nucleon separation energy
\cite{jan04}.
In the representation of Fig.~\ref{highem} the structure of the ridge stemming from two-nucleon knockout
is most significant. The results obtained for the \He4\eep\ reaction \cite{lee01} were presented
in the same way. Both data-sets exhibit qualitatively similar features.
Note that the cross sections systematically decrease at increasing \gammapq\
and that the values of \emiss\ at which the cross section reaches a maximum
increase with increasing \gammapq.
% Note that the cross sections systematically decrease at increasing \gammapq\ by about
% a factor three at $q = 300$ \mevc\ and a factor TWENTY [{\it in the previous version this was four!, LL}] at  $q = 400$ \mevc,
% and that the values of \emiss\ at which the cross section reaches a maximum
% increase with increasing \gammapq. 
This is characteristic for a
two-nucleon knockout reaction in which only one of the ejectiles is detected,
as expressed by Eq.~(\ref{eq:Em-ridge}).
In such a reaction the missing energy and momentum are largely accounted for
by the unobserved nucleon, as indicated by the dashed curve in Fig.~\ref{phase_space}.
Hence, the angles at which the cross sections reach the maximum  values and the widths
of the distributions are characteristic for the internal initial-state momentum of the
knocked-out nucleon pair.

The various curves in Fig.~\ref{highem} represent the results of
unfactorized distorted-wave calculations, performed for the two values
of the transferred momentum, i.e. 300 and 400 \mevc.  The dotted curve
accounts for the contribution of single-proton knockout from the $1s$
shell. For \gammapq $\geq$ 25\degree\ this cross section becomes
negligibly small.  The computed $(e,e'p)$ strength attributed to the
SRC, i.e.  central short-range and tensor correlations, are presented
by the dashed curves.  This strength is produced through the one-body
currents and would vanish identically in a mean-field theory. The
$(e,e'p)$ strength due to the genuine two-body currents,
i.e. meson-exchange and isobar currents, are presented by the dot-dash
curves. Finally, the solid curves represent the coherent sum of all
contributions. It is observed that the cross sections stemming from
the SRC (or, one-body currents) and those from meson-exchange and
isobar currents (or, two-body currents) exhibit a similar dependence
on \emiss.  In the considered kinematic domain the correlations
(dominated by the $NN$ tensor contribution) account for about a third
of the total strength.

Comparison of the data with the calculated cross sections indicates
that there is acceptable overall agreement.
The calculations  reasonably reproduce the dependence on \emiss\ of the cross sections at larger
values of \gammapq.
% At a fixed value of \gammapq\ the experimental cross sections decrease by a factor three to four
% between $q = 300$ \mevc\ and 400 \mevc. 
For both kinematics I and II, the experimental cross sections decrease by large factors, 
in the range between three to twenty, depending on the value of  \gammapq.
This reduction would be about
30\% if caused only by single-proton knockout, proportional to the proton electromagnetic form factors.

In Ref. \cite{liyan01} the theoretical model
presented here was compared to \O16\eep\ JLAB data that cover a
range of missing energies and momenta comparable to the present data.
However, the JLAB data were obtained in quasi-elastic kinematics at
considerably larger values of the energy and momentum transfer
(\omegaq) = (439 MeV, 1000 \mevc). The model could account for
the transverse nature and the shape of the JLAB data, but for only half of
the magnitude of the measured cross sections.

In  Ref. \cite{lee01} the  \He4\eep\ data taken at
${\omega} = 215$ MeV and  $q = 401$ \mevc\ are presented and compared
to the results of microscopic calculations performed by Laget. 
The experimental and calculated cross sections presented in Fig.~\ref{highem}
exhibit similar features as those for the \He4\eep\ reaction.
Indeed, the effect of SRC (one-body currents) gradually decreases at increasing \gammapq. At  \gammapq=80\degree\ meson-exchange and isobar currents were found
 to account for about 65\% of the cross section. A similar observation emerges
 from the  \O16\eep\ reaction.

In conclusion, the comparison of the present \O16\eep\ data with advanced model calculations shows that, at missing energies above about 25 MeV and in the studied kinematic range of missing momentum, single-proton knockout is manifestly small. In this energy region the contributions from two-body (meson-exchange and isobar) currents and to a lesser extent those from short-range correlations dominate the cross sections. In order to further distinguish between the latter two contributions in future experiments a longitudinal-transverse separation of the cross section is mandatory.

We would like to thank S.~Colilli, R.~Crateri, F.~Giuliani, M.~Gricia,
M.~Lucentini and F.~Santavenere for the preparation and maintenance 
of the waterfall target.

This work is part of the research program of the Foundation for Fundamental Research
of Matter (FOM), which is financially supported by the
Netherlands' Organisation for the Advancement of Pure Research (NWO).

\end{document}